\documentclass[12pt,a4paper]{article}              
\usepackage{graphicx}
\usepackage{xcolor}
\usepackage{amsmath}
\usepackage{authblk}
\usepackage{biblatex}
\bibliography{iucr}

\title{X-ray photon correlation spectroscopy of protein dynamics at nearly diffraction limited storage rings}

\author[1]{Johannes M\"oller} 
\author[2]{Michael Sprung}
\author[1]{Anders Madsen}
\author[3]{Christian Gutt}
\affil[1]{European X-ray Free Electron Laser Facility, Holzkoppel 4, Schenefeld, Germany}
\affil[2]{Deutsches Elektronen Synchrotron DESY, Hamburg, Germany}
\affil[3]{Department Physik, Universit\"at Siegen, D-57072 Siegen, Germany}

\begin{document}    
\maketitle                        

\begin{abstract}
This study explores the possibility to measure dynamics of proteins in solution using X-ray photon correlation spectroscopy (XPCS) at nearly diffraction limited storage rings (DLSR). We calculate the signal to noise ratio (SNR) of XPCS experiments from a concentrated lysozyme solution at the length scale of the hydrodynamic radius of the protein molecule. We take limitations given by the critical X-ray dose into account and find expressions for the SNR as a function of beam size, sample-detector distance and photon energy. Specifically, we show that the combined increase in coherent flux and coherence lengths at the DLSR PETRA IV will yield an increase in SNR of more than one order of magnitude. The resulting SNR values indicate that XPCS experiments of biological macromolecules on nm length scales will become feasible with the advent of a new generation of synchrotron sources. Our findings provide valuable input for the design and construction of future XPCS beamlines at DLSRs.
\end{abstract}


\section{Introduction}
Dynamics in concentrated protein systems are of fundamental interest in fields such as protein crystallization \cite{Durbin1996}, phase separation \cite{Anderson:2002aa}, the glass transition \cite{PhysRevLett.99.118301} or diffusion in crowded environments \cite{ELLIS2001597}, to name just a few. These systems display relatively slow and heterogeneous dynamics ranging from micro-seconds to seconds on length scales ranging from micrometers down to the single-particle nanometer scale. X-ray photon correlation spectroscopy (XPCS) is well suited to cover this length scale and time window employing coherent X-ray beams and tracing fluctuations in X-ray speckle patterns \cite{sutton_coherence,Gruebel2008,Perakis2017,Madsen2018}. 
However, the highly intense X-ray beams of synchrotron storage rings are also the cause of considerable radiation damage to the samples. Atomic scale XPCS experiments use X-ray doses of MGy and beyond, which can lead to beam induced dynamics, even in hard condensed matter samples \cite(Ruta2017). Soft and biological matter samples are much more sensitive to radiation damage requiring flowing samples \cite{Fluerasu:ot5581} \cite{Vodnala2018} or scanning samples with optimized data taking strategies \cite{Verwohlt2018}. Radiation damage of bio-molecules in solution is caused mainly by two effects: either via direct damage to the protein structure itself by photo-ionization or by indirect damage via hydrolysis of the surrounding water molecules (see e.g. \cite{Garman:ba5150}, \cite{Holton_A_2009}). In both cases, the damage becomes apparent by a characteristic change to the SAXS pattern indicating an increase of the radius of gyration, mostly due to aggregation. Typical critical X-ray doses for protein molecules in solution range from 7-10 kGy (BSA) to 0.3 kGy (Rnase) after which a degradation of the SAXS patterns become visible \cite{Jeffries_Limiting_2015}. These doses are easily reached within ms when using focused beams of modern synchrotron sources. While in protein crystallography cryogenic cooling helps to prevent the diffusion of radicals, such an approach is obviously impossible when studying the dynamics of proteins in solution. 

XPCS requires a coherent X-ray beam and the signal to noise ratio (SNR) in XPCS experiments ideally scales linear with the source brilliance $B$ \cite{Lumma2000}. The fastest accessible time scale then scales with $B^2$ promising four orders of magnitude faster temporal resolution at the upgraded sources of ESRF  and PETRA IV \cite{Einfeld:xe5006}\cite{Weckert:it5005} \cite{Schroer_2018} which is one of the key drivers for XPCS at DLSR sources \cite{Shpyrko_X_2014}. These arguments, however, only hold if radiation damage is no issue. Thus, the question arises of how much XPCS experiments of biological / radiation sensitive samples could really benefit from the gain in coherence performance of DLSR rings. Here, we show that the combination of (i) larger coherence lengths, (ii) higher photon energy and (iii) the increased coherent photon flux yields indeed an increase in SNR of up to one order of magnitude when compared to standard XPCS setups at today's storage rings. We calculate explicitly, using the boundary conditions set by the maximum tolerable X-ray doses of a  lysozyme solution, the XPCS speckle contrast, speckle intensities and maximum number of images per spot. We come to the conclusion, that DLSR rings hold the promise to measure dynamics of biological samples at length scales of a single protein molecule. \\

\section{XPCS on protein solutions}

XPCS experiments track fluctuations in X-ray speckle patterns yielding access to the intermediate scattering function $f(q,\tau ) = S(q,\tau)/S(q)$ by correlating intensities per detector pixel \cite{GRUBEL20043}.
The measured signal in such experiments is the normalized intensity autocorrelation function
\begin{equation}
g_2(q,\tau)=\frac{ \langle I_{pix}(q,t')I_{pix}(q,t'+\tau) \rangle }{\langle I_{pix}(q,t') \rangle ^2}=1+\beta |f(q,\tau)|^2,\label{eq:g2}
\end{equation}
with $\beta$ denoting the speckle contrast and $q = 4 \pi \sin (\Theta/2)/\lambda$ being the scattering vector, depending on the wavelength $\lambda$ and the scattering angle $\Theta$. The time delay between two consecutive time frames is denoted $\tau$ and $\left\langle \ldots \right\rangle$ is the ensemble average over all equivalent delay times $\tau$ and pixels within a certain range of the absolute value $\left|\vec{q}\right|$.\\

The scattering intensity per pixel from a protein solution is given by
\begin{equation}
I_{pix}(q)  = F_c \cdot t_{fr} \cdot T_{sample} \cdot d \cdot \frac{\mathrm{d}\Sigma}{\mathrm{d}\Omega}(q) \cdot \Delta \Omega_{pix},\label{eq:I}
\end{equation}
with $F_c$ denoting the incident coherent flux (ph/second), $t_{fr}$ the exposure time for one frame, $T_{sample}$ the sample's transmission and $\Delta \Omega_{pix} = (P / L)^2$ the solid angle covered by a single pixel, with $P$ being the pixel size and $L$ the sample-detector distance. In the following, we will set the thickness of the sample $d(E)$ to be equal to the absorption length of water $d(E) = 1/\mu(E)$ at each respective photon energy $E$, with the transmission following as $T_{sample} = \exp(- \mu d) \approx 0.368$. \\
The differential scattering cross section per unit volume or absolute scattering intensity in 1/m of a protein solution is defined as
\begin{equation}
\frac{\mathrm{d}\Sigma}{\mathrm{d}\Omega} (q) = C \cdot M \cdot \bar{v}^2 \cdot \Delta \rho^2 \cdot P(q) \cdot S_{eff}(q),
\end{equation}
with $P(q)$ the form and $S_{eff}(q)$ the effective structure factor and $C$ the protein concentration. We will calculate the SNR for lysozyme as model protein with a molar mass of $M=14.3$ kDa and specific volume $\bar{v} = 0.74$ cm${}^2$/g. The scattering contrast $\Delta \rho$ follows from the chemical composition of lysoszyme showing almost no dependence on energy in the energy range of interest here. With this, the absolute scattering intensity can be expressed as 
\begin{equation}
\frac{\mathrm{d}\Sigma}{\mathrm{d}\Omega} (q) = C \cdot 1.02 \mathrm{m^2 \over kg} \cdot P(q) \cdot S_{eff}(q),
\end{equation}
in good agreement with measured values of $(1.03 \pm 0.06) \mathrm{m^2 \over kg}$ ZITAT\cite{}.\\

\begin{figure}
\centering
		\includegraphics[width=1\textwidth]{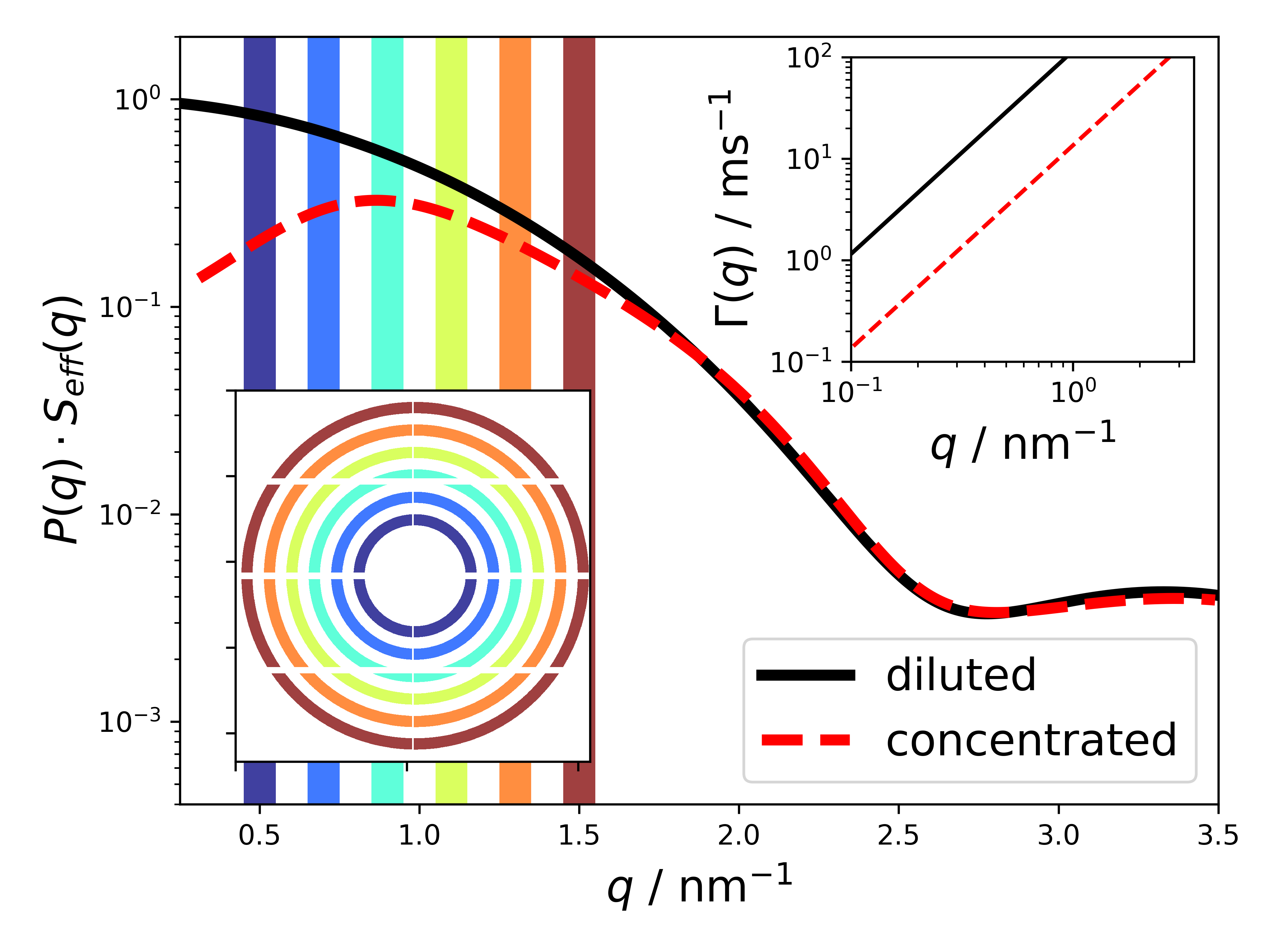}
	\caption{Form factor $P(Q)$ (black line) and effective structure factor $P(q) \cdot S_{eff}$ (red dashed line) of a diluted and concentrated lysozyme solution, respectively. The inset shows the relaxation rate $\Gamma(q)$ as a function of $q$ for both cases.}
	\label{fig:BioSAXS}
\end{figure}

 The form and effective structure factor $P(q) \cdot S_{eff}(q)$ are modeled following \cite{Moller:2012aa} and displayed in Fig. \ref{fig:BioSAXS} for a diluted ($10$ mg/ml) and concentrated ($250$ mg/ml) lysozyme solution. The $q$ values of interest are within $q = 0.5 \ \mathrm{nm}^{-1} - 1.5 \ \mathrm{nm}^{-1}$, which corresponds to length scales of $4-12$ nm.\\
 The dynamics of the low concentrated protein solution can be described as Brownian diffusion with a single exponential autocorrelation function
 \begin{equation}
     g_2(q,t)-1 = \beta \exp(-2\Gamma(q)t)
 \end{equation}
and relaxation rate
 \begin{equation}
     \Gamma = D_0 q^2
 \end{equation}
which is proportional to the Stokes-Einstein diffusion constant
 \begin{equation}
     D_0 = \frac{k_B T}{6 \pi \eta R_H},
 \end{equation}
 where $T$, $\eta$, $R_H$ and $k_B$ are the temperature, the viscosity of the suspending medium, the hydrodynamic radius of the protein and the Boltzmann constant, respectively. The $q$-dependence of the relaxation rate is plotted in the upper right inset of Fig. \ref{fig:BioSAXS} for diluted and concentrated lysozyme solutions. For the diluted case, we take the viscosity of water and a hydrodynamic radius of $R_H = 1.9$ nm. In order to illustrate the expected timescales for XPCS experiments on concentrated protein solutions we use an increased effective solution viscosity by a factor of 15 \cite{Godfrin2015}. The time scales of interest are here ranging from 100 $\mu s$ to seconds.\\
In practice, XPCS correlation functions are averaged over many pixels in a narrow range of $q$ values. Typical regions of interested are sketched as colored areas in Fig. \ref{fig:BioSAXS}. The same set of regions is additionally depicted in the lower left inset, showing the location of the corresponding pixels on an EIGER 4M detector for $E = 8$ keV and a sample to detector distance of $L=2$ m. In the following, we will always calculate the SNR at the maximum of the structure factor peak at $q=0.9 \ \mathrm{nm}^{-1}$.\\

\section{Signal to noise ratio}

The signal to noise ratio for the autocorrelation function $g_2(q,\tau)$ depends on the average intensity per pixel $I_{pix}$, the contrast $\beta$, the number of pixels $N_{pix}$, the number of frames $N_{fr}$ and the number of repetitions $N_{rep}$ via
\begin{equation}
SNR = \beta \cdot I_{pix} \cdot \sqrt{N},
\end{equation}
with $N = N_{pix} \cdot N_{fr} \cdot N_{rep}$.\\
Considering $N_{fr} = T/t_{fr}$ with $t_{fr}$ being the single frame exposure time and $T$ the total accumulated time for $N_{fr}$ frames yields in combination with Eq. \ref{eq:I} $SNR \propto F_c \sqrt{t_{fr} \cdot T}$. This scaling implies that an increase in coherent flux by one order of magnitude gives access to two orders of magnitude faster dynamics for the same SNR. However, this argument only holds when the sample is capable of handling the increased photon flux. If a critical dose $D_c$ exists, beyond which radiation induced damage starts to degrade the sample, the longest overall exposure time $T$ depends on $F_c$ and the increase of coherent flux might be less or not beneficial at all for studying radiations sensitive samples. 

The dose per second delivered to the sample depends on the photon flux as well as the photon energy which both also influences the achievable SNR. Here, we take all those parameters into account and calculate the benefit to the SNR from the increased coherent flux of DLSRs.We identify three parameter, which we will assume to be nearly free of choice over a wide range of values. These are the photon energy $E = \hbar c / \lambda$, the diameter $a$ of the X-ray beam spot size on the sample and the distance $L$ between sample and detector. In the following, we will establish the dependencies of the different contributions on the SNR, and determine the optimal set of $a$, $\lambda$, and $L$ values for XPCS experiment using radiation sensitive samples.\\

\begin{figure}
\centering
		\includegraphics[width=10cm]{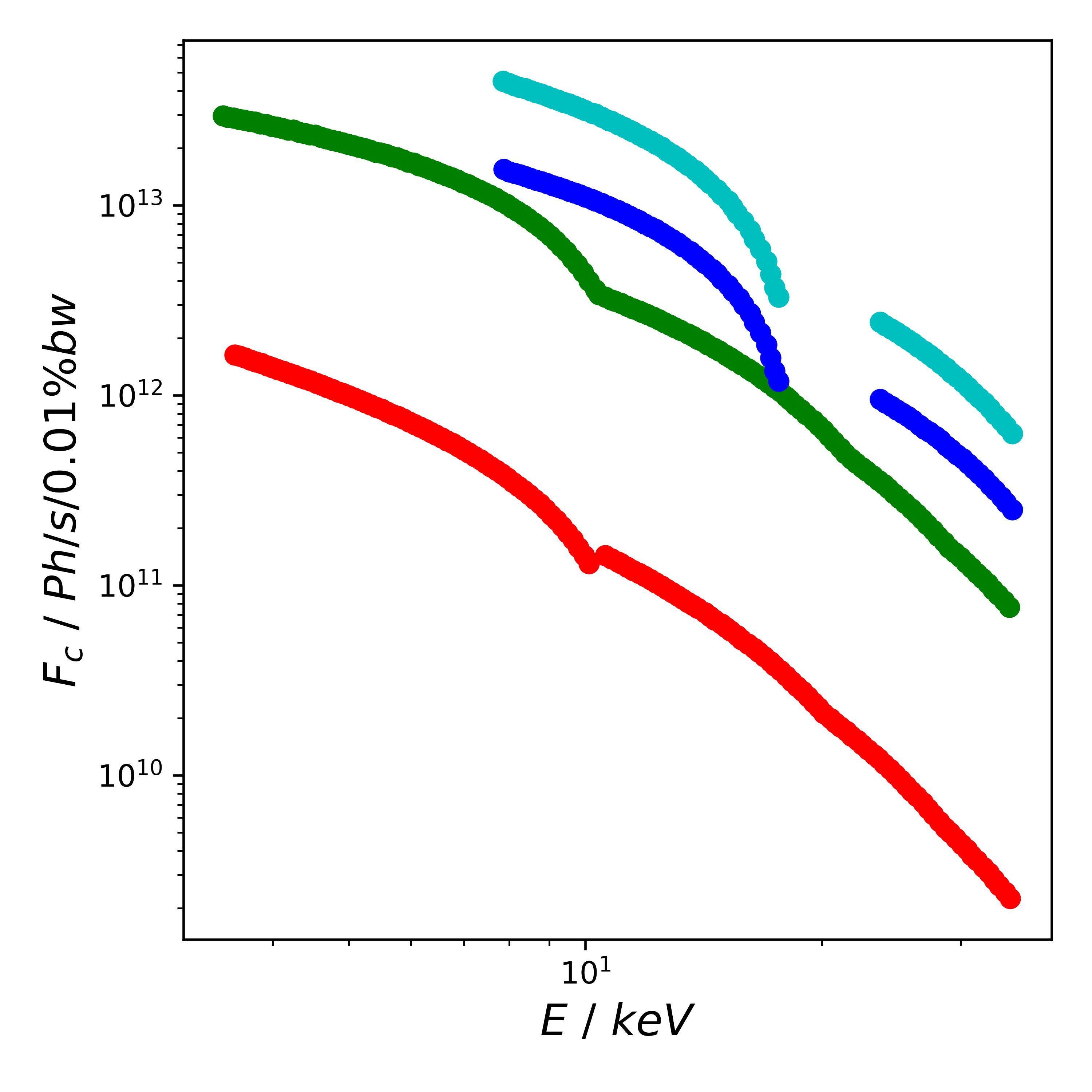}
	\caption{a) Brilliance taken from \cite{Schroer_2018} b) Coherent Flux calculated from fig. a), assuming a bandwidth of $0.01\%$, corresponding to a Si(111) monochromator with $\Delta \lambda / \lambda = 10^{-4}$.}
	\label{fig:bril}
\end{figure}

Fig. \ref{fig:bril} a) shows the expected increase of Brilliance as a function of photon energy for an U29 undulator at PETRA III and IV. Additionally, the case of an U18 with $5$ and $10$ m length will be investigated. The data shown is taken from \cite{Schroer_2018}. From this, the coherent flux can be calculated as 
\begin{equation}
F_c[ph / s / 0.1\%] = 10^{-8} Br[ph / s / 0.1\% / mm^2 / mr^2] \left(\lambda[\AA] \over 2\right)^2,
\end{equation}
which is also depicted in Fig. \ref{fig:bril} b). Using the given brilliance we calculate the coherent flux for $8$ keV at PETRA III as $3.8 \cdot 10^{11}$ ph/s. This is in good agreement with measured values of $2.3 \cdot 10^{11}$ ph/s, taking into account transmission effects of beamline components and optics. In the following, the actual coherent flux on the sample will be calculated by taking into account the same beamline transmission factor for all undulators.

\subsection{Limitations due to radiation damage}
We assume a critical dose $D_c$ beyond which radiation induced damage starts to degrade the sample which can be expressed as (Meisburger (2013). Biophys. J.  104, 227–236)
\begin{equation}
D_c = \frac{F_c E (1-T_{sample}) T}{d(E) a^2 \rho},
\end{equation}
with $F_c$ the photon flux on the sample, the product of energy dependent sample thickness $d(E)$ and beam area $a^2$, the sample absorption $(1-T_{sample})$, photon energy $E$ and exposure time $T$. From this we derive the maximum number of frames which can be measured before radiation damage occurs to be
\begin{equation}
 N_{fr} = \frac{d(E) a^2 \rho D_c}{t_{fr}  F_c E (1-T_{sample})}, \label{N}
\end{equation}
ignoring the latency time of the detector and absorption within the sample container walls. The sample thickness $d(E)$ is always adapted to the energy dependent absorption length of water. One important conclusion from equation \ref{N} is that the SNR scales via SNR $ \propto F_c \sqrt{N_{fr}} \propto \sqrt{F_c}$ for radiation sensitive samples.  Moreover, with the scalings $d(E) \propto E^3$ and $F_c \propto Br(E) /E^{2}$ we also find the peculiar relation of $N_{fr} \propto E^4$ favoring higher photon energies if a  large number of frames is required. \\
We illustrate this with the example of a typical spot size for XPCS experiments of $a = 4 \ \mathrm{\mu m}$, an exposure time of a single frame of $t_{fr} = 1$ ms and a critical Dose limit for a concentrated lysozyme solution of $D_c = 1$ kGy. 
\begin{figure}
\centering
		\includegraphics[width=10cm]{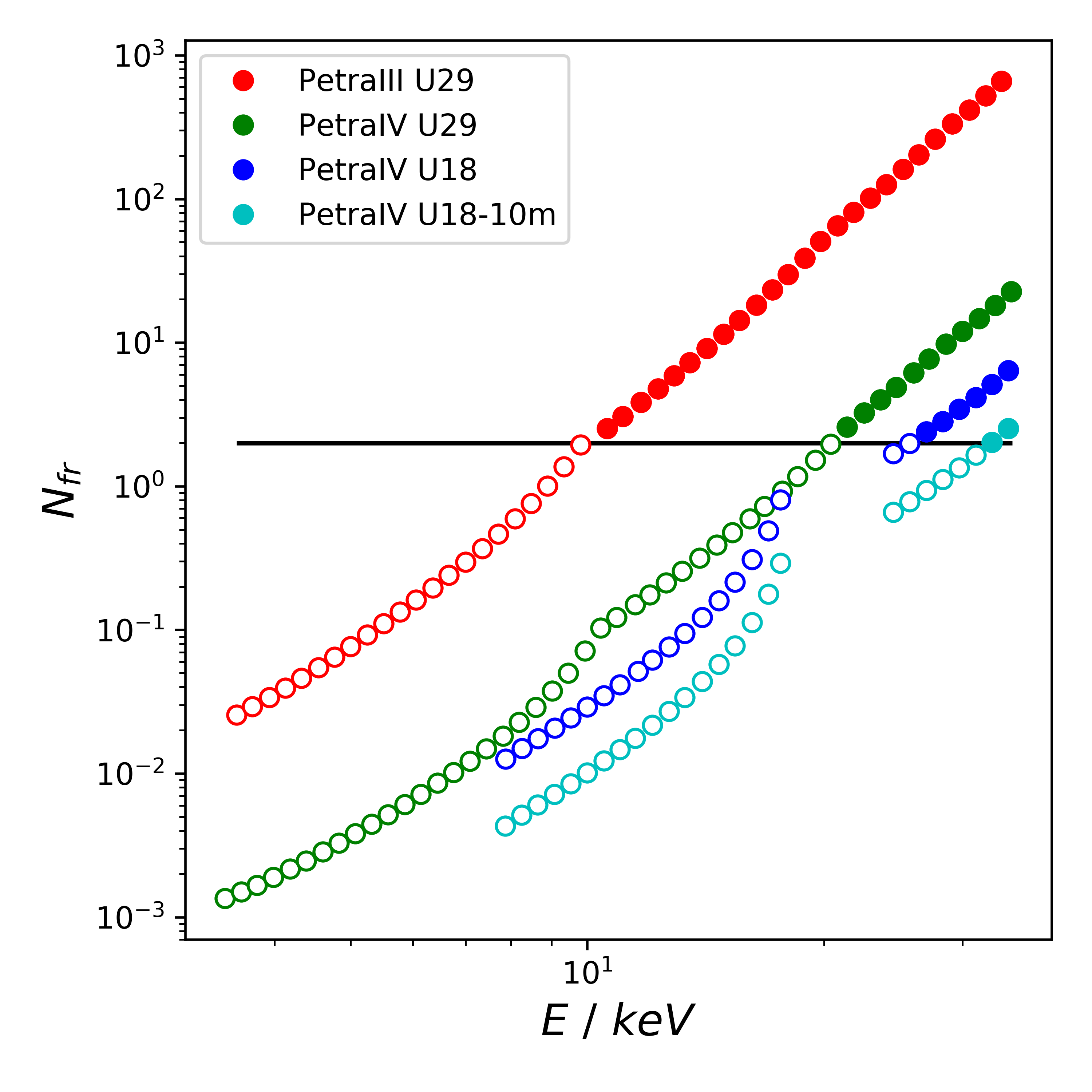}
	\caption{Maximum number of frames which can be measured on one spot before the onset of radiation damage for a lysozyme solution with beamsize $a = 4 \ \mathrm{\mu m}$ and exposure time per frame of 1 ms. The vertical black line depicts the threshold of at least two consecutive frames.}
	\label{fig:Nfr}
\end{figure}
Fig. \ref{fig:Nfr} displays the possible number of consecutive frames as a function of photon energy.  A prerequisite for correlation spectroscopy is obviously that the number of consecutive frames is at least two (i.e. $N_{fr} \geq 2$), indicated via filled symbols. Already with the coherent flux of PETRA III the critical dose is exceeded after or during the first image and beam damage is occurring between two images for photon energies below $10$ keV. At this energy, an increase in coherent flux would therefore not be usable for XPCS experiments on protein samples. However, it can also be seen that $N_{fr}$ increases with photon energy due to the increasing absorption length of the X-rays. Effectively, the radiation dose is spread over a larger sample volume with increasing photon energy. However, many properties like the speckle size, the coherent flux as well as the longitudinal and transversal coherence lengths decrease with increasing photon energy. Therefore, the disadvantageous influence of these properties on the speckle contrast $\beta$ and consequently on the SNR of XPCS experiments need to be taken into account as well.

\subsection{Speckle contrast $\beta$}
\begin{figure}
\centering
		\includegraphics[width=10cm]{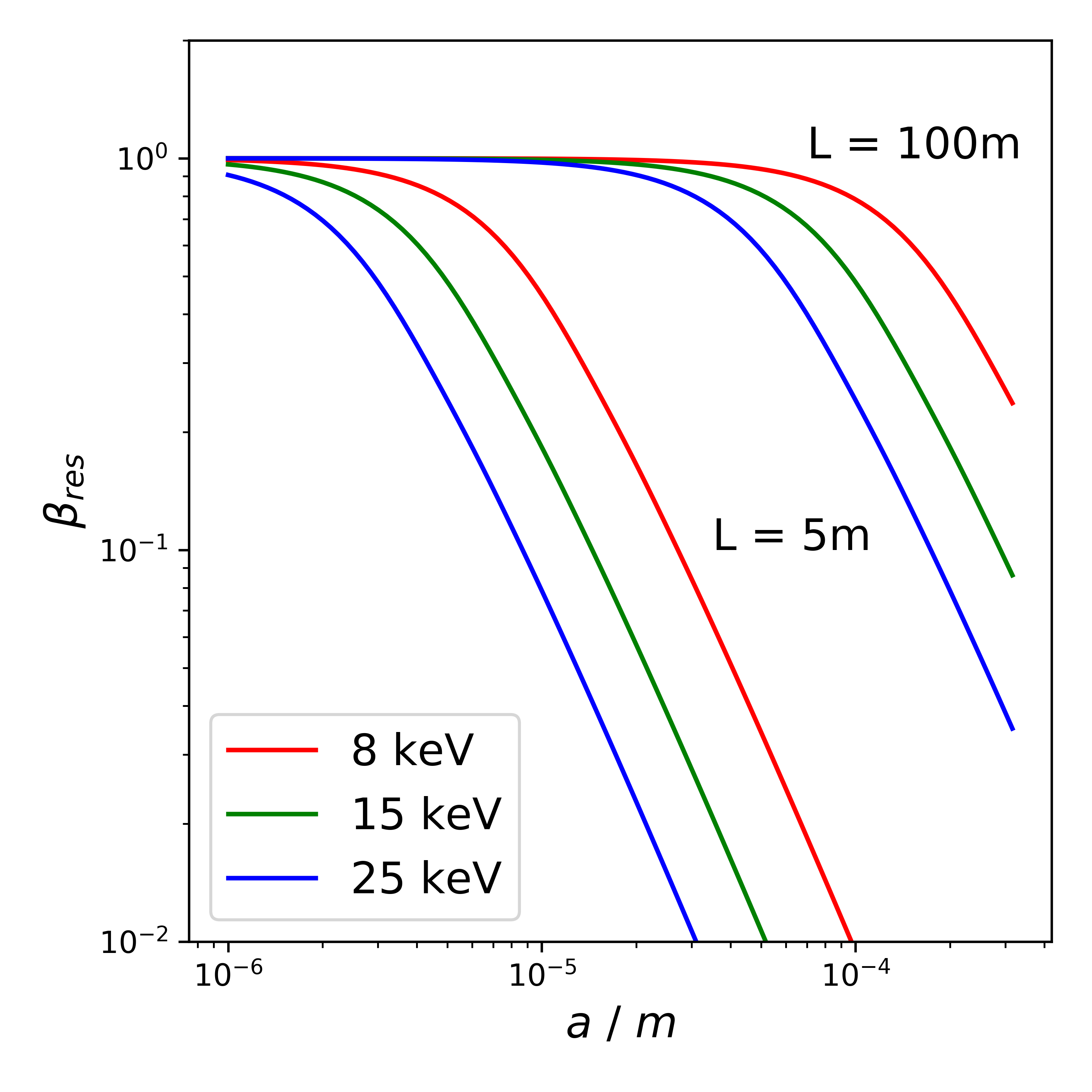}
	\caption{Speckle contrast $\beta_{res}$ as a function of beam size $a$ on the sample according to eq. \ref{eq:betares}. Calculated for photon energies of 8 keV (red), 15 keV (green) and 24 keV (blue) and for sample-detector distances L=100 m (top lines) and L=5 m (bottom lines), respectively.}
	\label{fig:betares}
\end{figure}

The speckle contrast depends on nearly all experimental parameters such as pixel size $P$, speckle size $S\approx \lambda L/a$, beam size $a$, sample thickness $d$, wavevector transfer $q$, and the transverse and longitudinal coherence lengths. It can be written as a product, 
\begin{equation}
\beta(a, d, q, \lambda, L) = \beta_{cl}(a,d,q,\lambda) \beta_{res}(a,L,\lambda)
\end{equation}
in which the first factor $\beta_{cl}$ corresponds to the reduction of the contrast from unity due to the finite coherence lengths in transverse and longitudinal direction. The second factor $\beta_{res}$ corresponds to a finite angular resolution of the experimental setup. This results in a reduction of contrast if the pixel size of the detector $P$ exceeds the size of the speckle $S$:
\begin{equation}
\beta_{res}(a,L,\lambda) = \left( {2 \over w^2} \int_0^w (w-v) \left( \frac{\sin(v/2)}{v/2}\right)^2 \mathrm{d}v\right)^2 \label{eq:betares} 
\end{equation}
with $w= 2 \pi P a/L\lambda = 2 \pi P/S$. Fig. \ref{fig:betares} displays the speckle contrast $\beta_{res}$ as a function of beamsize $a$ for sample-detector distances of $L=5$ m and $L=100$ m, respectively, pixel size $P=75 \ \mu$m  and photon energies of 8, 15 and 25 keV.  The maximum $\beta_{res}$ is obtained in a high resolution configuration with $S \geq P$ and scales as $\beta \approx \lambda^2 L^2 /a^2P^2$ in the low resolution configuration, when $S \ll  P$. Therefore, XPCS experiments with large beamsizes require long sample-detector distances in order to resolve the smaller speckles.  

The dependence of $\beta_{cl}$ on beamsize $a$, sample thickness $d$, transverse $\xi_h$, bandwidth $\Delta \lambda \over \lambda$ and q-value is taken into account via {\cite{sutton_coherence}
\begin{equation} \label{eq:betacl}
\begin{split}
\beta_{cl}(a,d,q,\lambda) & = \frac{2}{(a \cdot d)^2} \int_0^{a} \mathrm {d}x \int_0^d \mathrm{d}z (a-x)(d-z) \exp(-x^2/\xi_h^2) \\
 &  \times \left( \exp(-2\left|Ax+Bz\right|)+\exp(-2\left|Ax-Bz\right|) \right), 
\end{split}
\end{equation}
with $A = { \Delta \lambda \over \lambda} q \sqrt{1- {1 \over 4} q^2 / k^2}$ and $B = - {\Delta \lambda \over 2 \lambda}{q^2 \over k}$. In vertical direction we assume a completely coherent beam and in horizontal direction, the coherence length is estimated as 
\begin{equation}
\xi_h = {R \cdot \lambda \over 2\pi \sigma} , \label{eq:xih}
\end{equation}
with $R$ being the distance between source and beam defining aperture and $\sigma$ the RMS source size. With $\sigma_h = 36 \ \mu$m (P10, low-$\beta$ source, 10 keV, $R = 90$ m), this results in a horizontal coherence length at $E=10$ keV of $\xi_h = 49 \ \mu$m. A reduced horizontal source size at PETRA IV of $\sigma_h = 12 \ \mu$m would result in an increased horizontal coherence length of $\xi_h = 147 \ \mu$m at the same energy. These values reduce to $20 \ \mu$m and $59 \ \mu$m at an energy of $E=25$ keV, respectively. The full energy dependence of $\xi_h$ is shown in Fig. \ref{fig:betacl} a).\\
Using a partially coherent source like a undulator for coherent scattering experiments, cutting of the incident X-ray beam is required in order to obtain a nearly fully transversely coherent beam. Therefore, a beam defining aperture is set to an opening size equal to the transversal coherence length. Smaller beam sizes can be achieved with additional focussing elements. For our calculations, we will consider the resulting focussed beam as fully coherent with a $\xi_h$ being equal to the beam size. For larger beamsizes, $\xi_h$ is calculated following Eq. \ref{eq:xih}.\\
The temporal or longitudinal coherence length can be calculated as
\begin{equation}
\xi_l = {\lambda \over 2} {\lambda \over \Delta \lambda},
\end{equation}
depending on the bandwidth of the used monochromator ( $\Delta \lambda / \lambda \approx 1.4 \cdot 10^{-4}$ for a Si(111) monochromator and $\Delta \lambda / \lambda \approx 3 \cdot 10^{-5}$ for Si(311)). \\
The results for $\beta_{cl}$ as a function of beam size and X-ray energy are shown in figure \ref{fig:betacl} b) for a q-value of $q=0.9 $ nm$^{-1}$, corresponding to the peak of the structure factor shown in Fig. \ref {fig:BioSAXS}.
\begin{figure}
\centering
		\includegraphics[width=10cm]{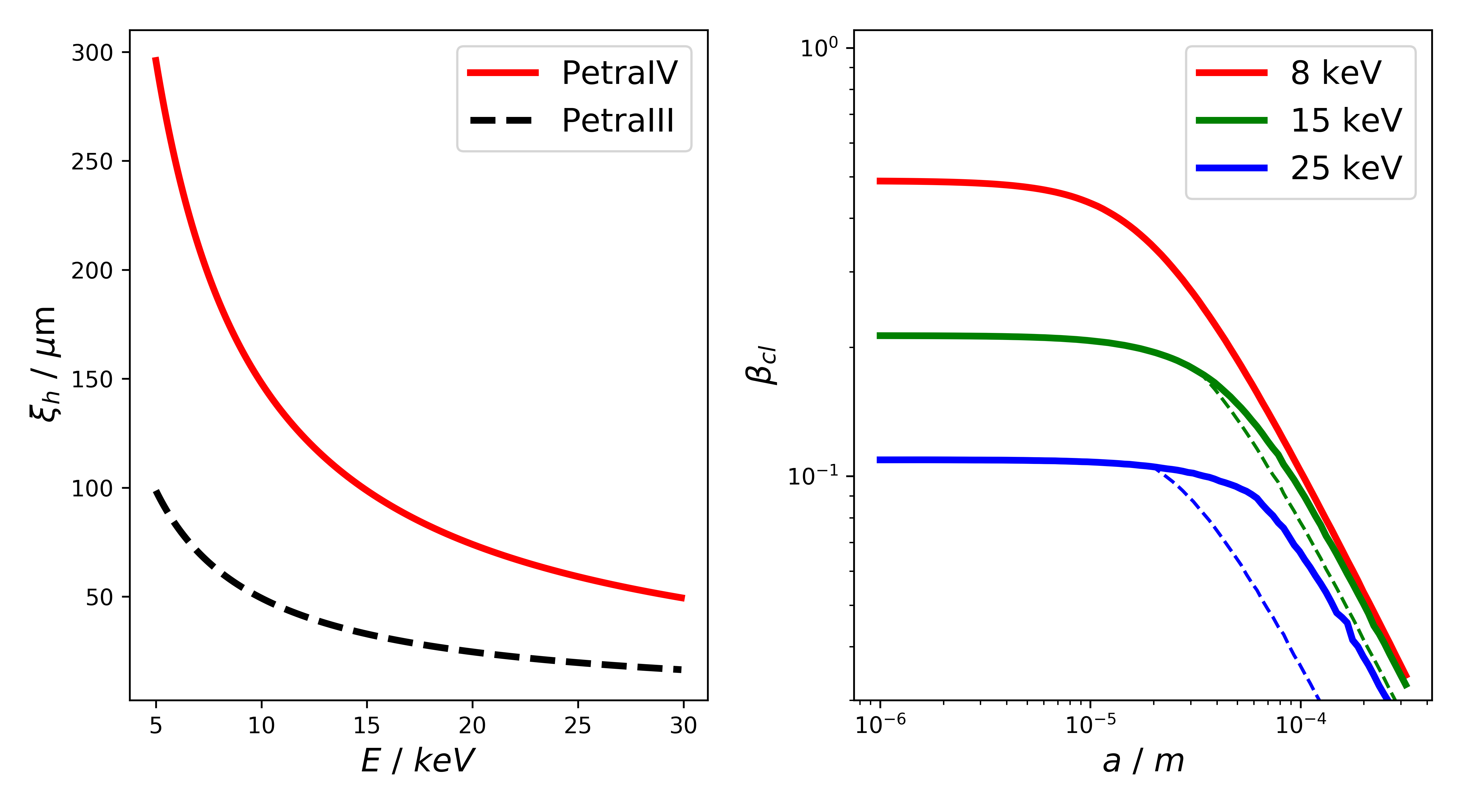}
	\caption{a) Horizontal coherence length calculated from the source properties of PETRA III and PETRA IV as a function of photon energy. b) Speckle contrast $\beta_{cl}$ as a function of beam size calculated according to eq. \ref{eq:betacl} (photon energies 8 keV (red), 15 keV (green) and 25 keV (blue)). The dashed line corresponds to the horizontal coherence length at P10 PETRA III, the solid line represents the horizontal coherence length expected with PETRA IV. The q-value is $q=0.9$ nm$^{-1}$ and the sample thicknesses are $d$=1.0, 6.5 and 23 mm corresponding to the absorption length of water at the respective photon energies.}
	\label{fig:betacl}
\end{figure}
We observe a reduction of speckle contrast with increasing beamsize and a reduced contrast for smaller beamsizes as a function of photon energy. Both reductions can be explained by the scattering volume, defined by spot size $a$ and sample thickness $d$, exceeding the coherence volume defined by the longitudinal and transversal coherence length. 

\subsection{Number of pixel}
Changing the photon energy and sample detector distance has direct implications on the number of pixels which can be covered within an area of a certain $q$-range. The scattering signal may be in a circular region of interest on the detector of width $\Delta q$ and radius $q$. In the SAXS regime, $q = (4 \pi/  \lambda) \theta $ and $\Delta q = (4 \pi/  \lambda) \Delta \theta$, and the diffraction ring has a width on the detector of $\Delta \theta \cdot L$ and a circumference of $2\pi (2\theta) L$. The number of illuminated pixels is thus
\begin{equation}
N_{pix} = \frac{q \Delta q \lambda^2 L^2}{4 \pi P^2}.
\end{equation}

\section{XPCS of protein solutions}

Having established the dependence of the SNR on the experimental parameters we can use the expression
\begin{equation}
SNR = \beta(a, \lambda, L) \cdot I_{pix}(\lambda, L)  \cdot \sqrt{N_{fr}(a, \lambda)\cdot N_{pix}(\lambda, L)}
\end{equation}
to characterize the influence of the improved brilliance of the new generation of X-ray sources on XPCS experiments with radiation sensitive samples.\\
In Fig. \ref{fig:StandXPCS}, we display the SNR for a standard XPCS setup. It was assumed that an EIGER 4M detector  \cite{Johnson2014} is used, with a sample detector distance of $L=5$ m, which corresponds at a photon energy of $E=8$ keV to the inset of Fig. \ref{fig:BioSAXS}. In order to match the speckle size to the pixel size, an X-ray spot size of $a=4$ $\mu$m is required, corresponding to the calculations shown in Fig. \ref {fig:Nfr}.\\
Further parameters are:
\begin{table}
\caption{Parameters fixed for the calculations of the SNR}
\begin{tabular}{ll}  
$q $&$ 0.9 \mathrm{nm}^{-1}$\\
$\Delta q $&$ 0.1 \mathrm{nm}^{-1}$\\
$C $&$ 250 \mathrm{mg/ml}$\\
$P(q) \cdot S(q) $&$ \approx 0.3$\\
$D_c $&$ 1,000 \mathrm{J/kg} = 1 \mathrm{kGy}$\\
$P $&$ 75 \mathrm{\mu m}$\\
$t $&$ 1 \mathrm{ms}$\\
\end{tabular}
\end{table}

\begin{figure}
\centering
		\includegraphics[width=10cm]{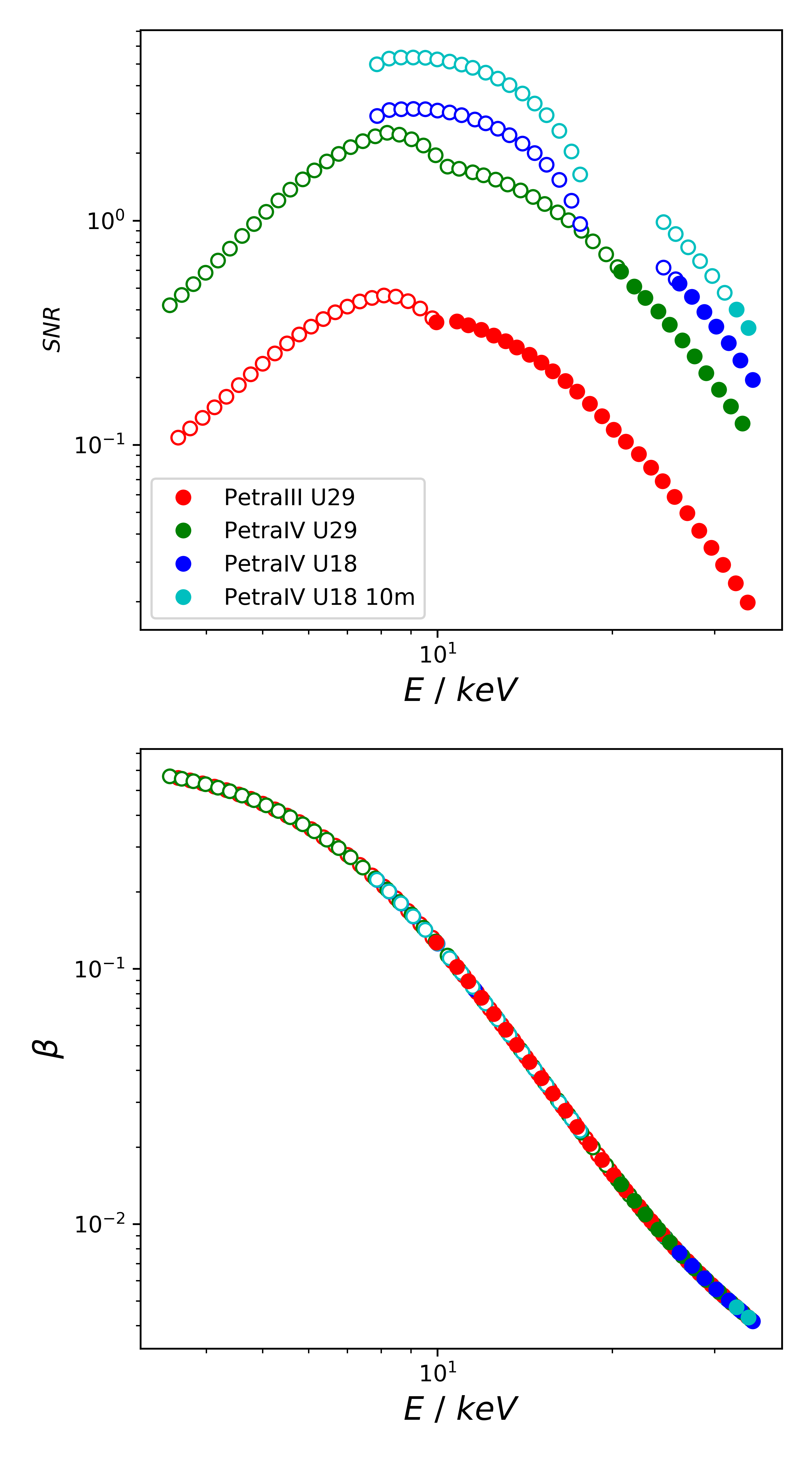}
	\caption{a) Signal to Noise ratio (SNR) calculated as a function of photon energy for a setup with ($a= 4 \mu$m, $L=2$ m) and for different undulators. Open symbols correspond to experimental conditions which are not accesible due to beam damage effects. b) Speckle contrast of this setup as a function of photon energy.}
	\label{fig:StandXPCS}
\end{figure}

The red data points correspond to the photon beam properties of PETRA III, the green, blue and cyan points to the improved coherent flux $F_c$ offered by PETRA IV with different undulators. As can be seen, the increasing coherent flux offers theoretically improved SNR values of more than one order of magnitude. However, as marked with open symbols, the highest theoretically possible SNR of each configuration corresponds to experimental conditions where the critical dose limit of the sample is reached within two sequential acquisitions (i.e. $N_{fr} \leq 2)$. Therefore, the maximum increase in SNR can not be reached in practice and the upgrade to PETRA IV would not lead to such a significant increase in SNR for this setup.\\
Data points which correspond to beam conditions where at least two sequential acquisitions are possible are displayed as filled symbols. It is evident that higher beam energies with also thicker samples would ease the effect of a higher flux and make XPCS experiments possible also with a standard configuration ($L=5$ m, $a=4 \ \mu$m). However, as displayed in Fig. \ref{fig:StandXPCS} b), this also results in much reduced speckle contrasts and therefore the beneficial effect of an increased coherent flux on the SNR is largely lost due to the strongly reduced speckle contrast $\beta$. 

\subsection{Optimizing the experimental setup}
In order to use the increased coherent flux for XPCS experiments, one has to adapt the experimental setup in terms of focussing, photon energy and sample detector distance.\\
Therefore, we repeat the previously presented calculations for a set of different beamsizes $a$ and sample-detector distances $L$. At each point in the $a-L$ plane, the SNR is calculated as a function of photon energy and the maximum is calculated. However, only values are considered which correspond to $N_{fr} \geq 2$ at 1 ms exposure. The maximum SNR for each pair of $a$ and $L$ values is displayed in Fig. \ref{fig:contour}.\\
It can be seen that the previously discussed setup with a small beam and large speckle (marked by a red dot) does not give the best SNR already for the case of PETRA III. With a sample-detector distance of $L=5.5$ m and an X-ray spot size of $a=9 \ \mu$m, the expected SNR increases by $25 \%$.\\
\begin{figure}
\centering
		\includegraphics[width=10cm]{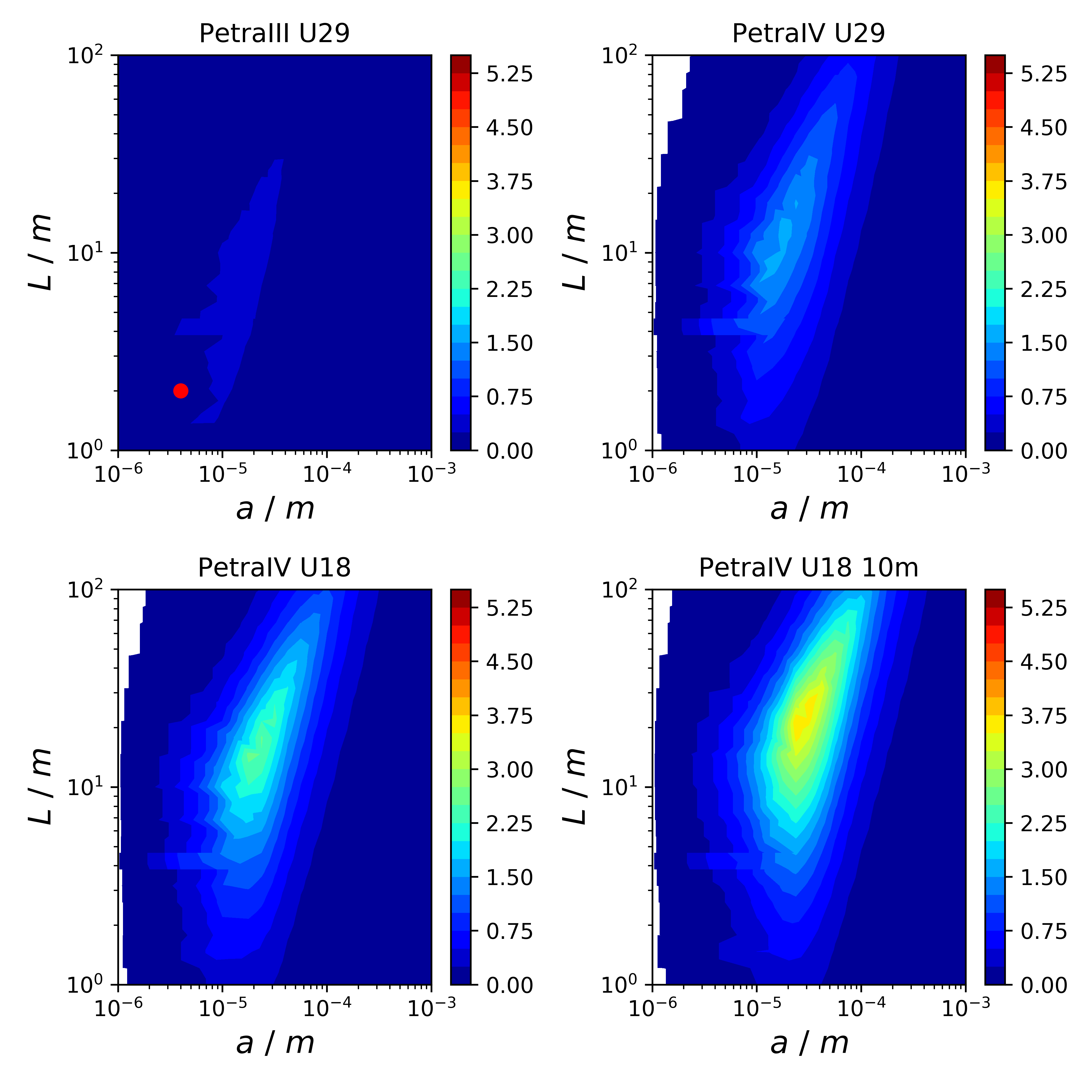}
	\caption{The maximum value of SNR as a function of sample-detector distance $L$ and beamsize $a$ for DLSR. The highest SNRs are $SNR_{P3}=0.46$; $SNR_{P4}=1.7$; $SNR_{P4 U18}=2.7$; $SNR_{P4 U18-10m}=3.9$ }
	\label{fig:contour}
\end{figure}
However, in the case of PETRA IV (U18-10m), an overall increase in SNR by about one order of magnitude can be achieved, without exceeding the critical radiation dose of the sample. This setup would feature a sample detector distance of $L=26 \ \mathrm{m}$ and a spot size of $a=24 \ \mathrm{\mu m}$ at $E=14.7$ keV.\\

\begin{table}
\caption{Parameters optimized setup for $N_{fr}=2$ using a Si(111) monochromator \label{tab2}}
\begin{tabular}{lllll}  
paramters &U29 (PIII)               & U29 & U18 5m & U18 10m \\
\hline
signal to noise ratio SNR           & $0.4$ &$1.7$& $2.6$ &$3.7$  \\
\hline 
beam size $a$ / $\mu$m              & $7.5$ &$13.3$ & $17.7$ &$23.7$    \\
\hline
sample detector distance $L$ / m    & $3.8$ &$10.0$ & $14.7$ &$21.5$ \\
\hline 
beam energy $E$ / keV               & $8.1$ &$12.2$& $13.6$ &$14.7$  \\
\hline
coherent Flux $F_c$ / ph/s          & $2.1  \ 10^{11}$ & $1.5  \ 10^{12}$ &$3.4 \ 10^{12}$ &$7.2 \ 10^{12}$  \\
\hline
contrast $\beta$                    & $0.20$ & $0.12$ & $0.10$ & $0.08$  \\
\hline
speckle size $S$  / $\mu$m          & $78$& $76$ & $75$ &$76$  \\
\hline
intensity per pixel $I_{pix}$ / ph/ms& $2.3 \ 10^{-3}$& $8.8.0 \ 10^{-3}$ & $1.2 \ 10^{-2}$ & $1.5 \ 10^{-2}$ \\
\hline
number of pixel in q-range $N_{pix}$ & 0.4 M& 1.3 M& 2.2 M& 4.2 M \\
\hline
number of frames $N_{fr}$           &2 & 2& 2 & 2 \\
\hline
sample thickness $d$ / mm           &$1.0$& $3.6$ & $4.9$ & $6.3$\\
\hline
exposed sample volume / nL          & $0.05$& $0.6$&  $1.6$ & $3.5$ \\
\end{tabular}
\end{table}

The resulting parameter for the optimized experimental setups are summarized in Tab. \ref{tab2} for each of the considered undulators. We note that for higher coherent flux setups the optimized setups feature an increase of  beam size $a$, sample detector distance $L$ and photon energy $E$. \\

As a general trend, it is evident that the sample volume, spanned by the sample thickness $d$ and spot size $a$ needs to be increased when the coherent flux increases. In order to compensate for the consequently decreasing angular speckle size, the sample detector distance needs to increase so that the speckle size can maintain its value of $S \approx 75$ $\mu$m. However, it can be seen that one can still observe a decrease in speckle contrast $\beta$, even though the speckle have the same size on the detector for all four presented setups. This effect is due to the second contribution to the speckle contrast $\beta_{cl}$, see Eq. \ref{eq:betacl}, originating from the limited longitudinal coherence length of the X-ray beam.

\subsection{Si(311)-Mono}
Here, we investigate how an additional increase of the longitudinal coherence length by using a Si(311) monochromator benefit the achievable SNR. We repeat the calculations with a reduced bandwidth of $3 \cdot 10^{-5}$ and a reduced flux compared to the Si(111) calculations by $74 \ \%$. The resulting SNRs are displayed in Fig. \ref{fig:contour_311} and Tab.‚ \ref{tab3}. We find that the use of a Si(311) monochromator improves the SNR by an additional $30 \%$ compared to the Si(111) thus leading to an overall SNR gain of a factor of 13 when comparing PETRA III with PETRA IV.

\begin{figure}
\centering
		\includegraphics[width=10cm]{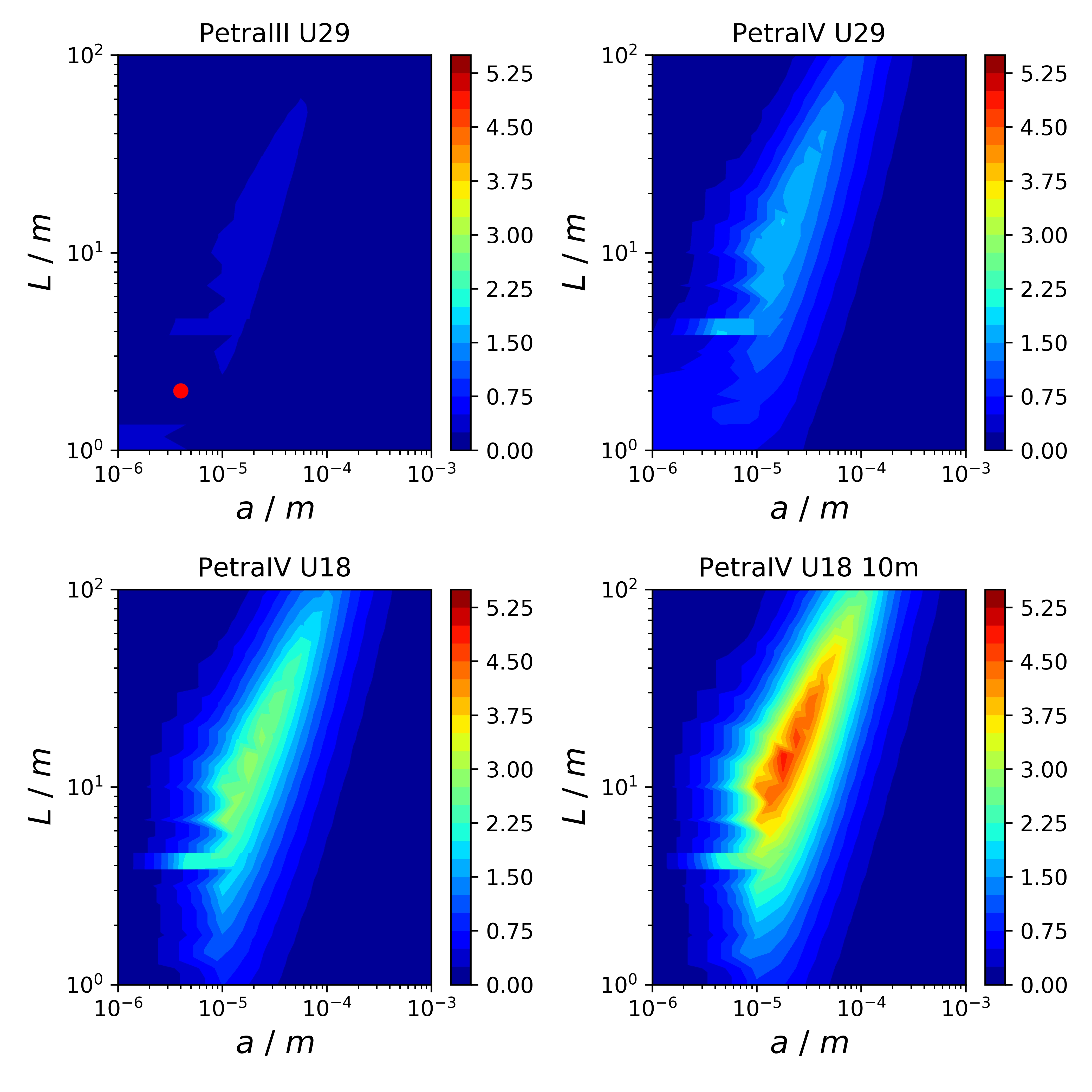}
	\caption{Best combination of sample-detector distance $L$ and beamsize $a$ for DLSR using an Si(311) monochromator.}
	\label{fig:contour_311}
\end{figure}

\begin{table}
\caption{optimized setup using a Si(311) monochromator \label{tab3}}
\begin{tabular}{lllll}  
parameters &U29 (PIII)               & U29 & U18 5m & U18 10m \\
\hline
signal to noise ratio SNR           & $0.37$ &$1.8$& $3.0$ &$4.9$  \\
\hline 
beam size $a$ / $\mu$m              & $7.5$ &$10.0$ & $13.3$ &$17.8$    \\
\hline
sample detector distance $L$ / m    & $5.6$ &$8.3$ & $10.0$ &$14.7$ \\
\hline 
beam energy $E$ / keV               & $12.5$ &$14.1$& $12.5$ &$13.8$  \\
\hline
coherent Flux $F_c$ / ph/s          & $1.5  \ 10^{10}$ & $2.9  \ 10^{11}$ &$1.1 \ 10^{12}$ &$2.3 \ 10^{12}$  \\
\hline
contrast $\beta$                    & $0.23$ & $0.20$ & $0.23$ & $0.20$  \\
\hline
speckle size $S$  / $\mu$m          & $74$& $73$ & $74$ &$74$  \\
\hline
intensity per pixel $I_{pix}$ / ph/ms& $3.0 \ 10^{-3}$& $3.9 \ 10^{-3}$ & $6.9 \ 10^{-3}$ & $9.1 \ 10^{-3}$ \\
\hline
number of pixel in q-range $N_{pix}$ & 0.06 M& 0.7 M& 1.3 M& 2.2 M \\
\hline
number of frames $N_{fr}$           &69 & 8& 3 & 3 \\
\hline
sample thickness $d$ / mm           &$3.8$& $5.5$ & $3.8$ & $5.2$\\
\hline
exposed sample volume / nL          & $0.21$& $0.55$&  $0.7$ & $1.6$ \\
\end{tabular}
\end{table}

\subsection{Multiple frame XPCS and two time correlation functions}
It becomes evident that with SNR values $\propto 3-5$ XPCS from protein solutions is indeed possible at DLSRs with adapted experimental setups. As a direct consequence of the presented results, the optimized data acquisition scheme differs from conventional XPCS measurements. Instead of taking many hundreds to thousands of images at one spot, the scheme with maximum SNR for protein XPCS rather consists of "double-shot" exposures. This would not give a full correlation function from one spot on the sample, but rather one data point of $g_2$ for each illuminated sample spot. In consequence, the correlation function would be constructed from many of such double-shot exposures, which each can be done on a new sample spot and with a different delay time $\tau$ between the two frames (see e.g. \cite{Verwohlt2018}). The required sample volume therefore scales with the desired number of data points of $g_2$.\\
However, this acquisition scheme is not suitable for samples displaying heterogeneous dynamics or aging effects. In such cases a movie-mode acquisition scheme with more than two frames per spot is needed. Fig. \ref{fig:contour_tt} displays the resulting SNR values in the $a-L$ plane for $N_{fr}=2, 5, 25, 100$ for the case of PETRA IV U18 10m.
We find that with increasing number of images $N_{fr}$ the value of the maximum SNR decreases and its position in the $a-L$ plane shifts towards larger beamsizes $a$ and larger sample-detector distances $L$. For realizing the higher number of frames, an increase of the photon energy and of the beamsize is required (from Eq. 11 we find the scaling $a \propto \sqrt{N_{fr}}$). The resulting degradation of speckle contrast is partially counterbalanced by improving the angular resolution via a larger sample-detector distance. For example for $N=100$ frames the optimum SNR is 3.2 at $a=75 \mu m$, $L=82$m and $E=15.4$ keV. Generally speaking we find  at the maximum of the SNR a scaling of $L\propto a \propto \sqrt{N_{fr}B(E)}$.

\begin{figure}
\centering
		\includegraphics[width=10cm]{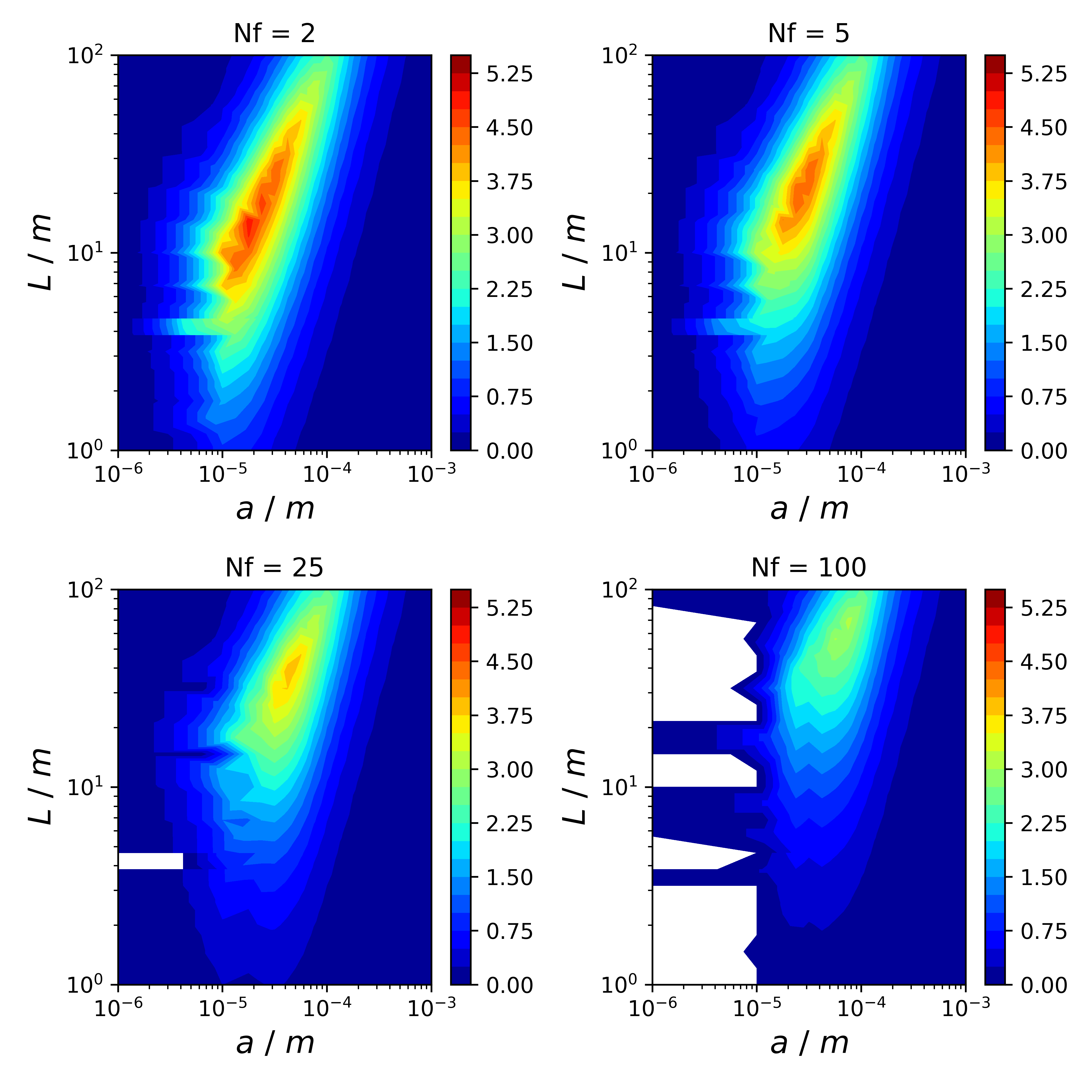}
	\caption{The maximum value of the SNR as a function of sample-detector distance $L$ and beamsize $a$ for PETRA IV (U18 10m), for $N_{fr}=2, 5, 25, 100$ number of frames taken on the same sample spot (Concentrated lysozyme solution, q=1 nm$^{-1}$ and critical dose of 1 kGy). The white color indicates $a-L$ combinations in which the required number of frames cannot be realized with the photon flux.}
	\label{fig:contour_tt}
\end{figure}

In reality it might become difficult to realize a beamline with up to $100$ m sample-detector distance, which consequently would also require a detector with a very large number of pixels. However, it can be seen from Fig. 9 that also at shorter sample-detector distances $L$, the SNR is still significantly larger than 1 for up to $N_{fr}=100$. Therefore, we investigate how the SNR can be optimized, if the length of the beamline is limited to a fixed value of $L$ and on the same time a certain number of frames $N_{fr}$ is required to track the physics of the protein solution.

We demonstrate this by fixing the sample-detector distance at $L=30$ m and using a Si(311) monochromator. We plot both the SNR and the maximum number of frames possible as a function of beamsize $a$ (Fig. 10 left) for photon energies of $E=13.1$ keV (solid), $E=14.9$ keV (dashed) and $E=17$ keV (dash-dotted), respectively. Fig. 10 right displays the SNR as a function of $N_{fr}$ for the different photon energies. The benefit of using slightly higher photon energies than 13 keV is obvious as it allows to either increase the SNR value at fixed $N_{fr}$ or to record more images at fixed value of the SNR.
We find that with the source parameters of PETRA IV (U18 10m) the resulting values of the SNR are on the order of single digits. Specifically, we may take the example of $N_{fr}=100$ and find an SNR value of 2.5. Thus with 100 repeats (i.e. $N_{rep}=100)$ we could obtain an SNR of 25 of an averaged correlation function.

\begin{figure}
\centering
		\includegraphics[width=10cm]{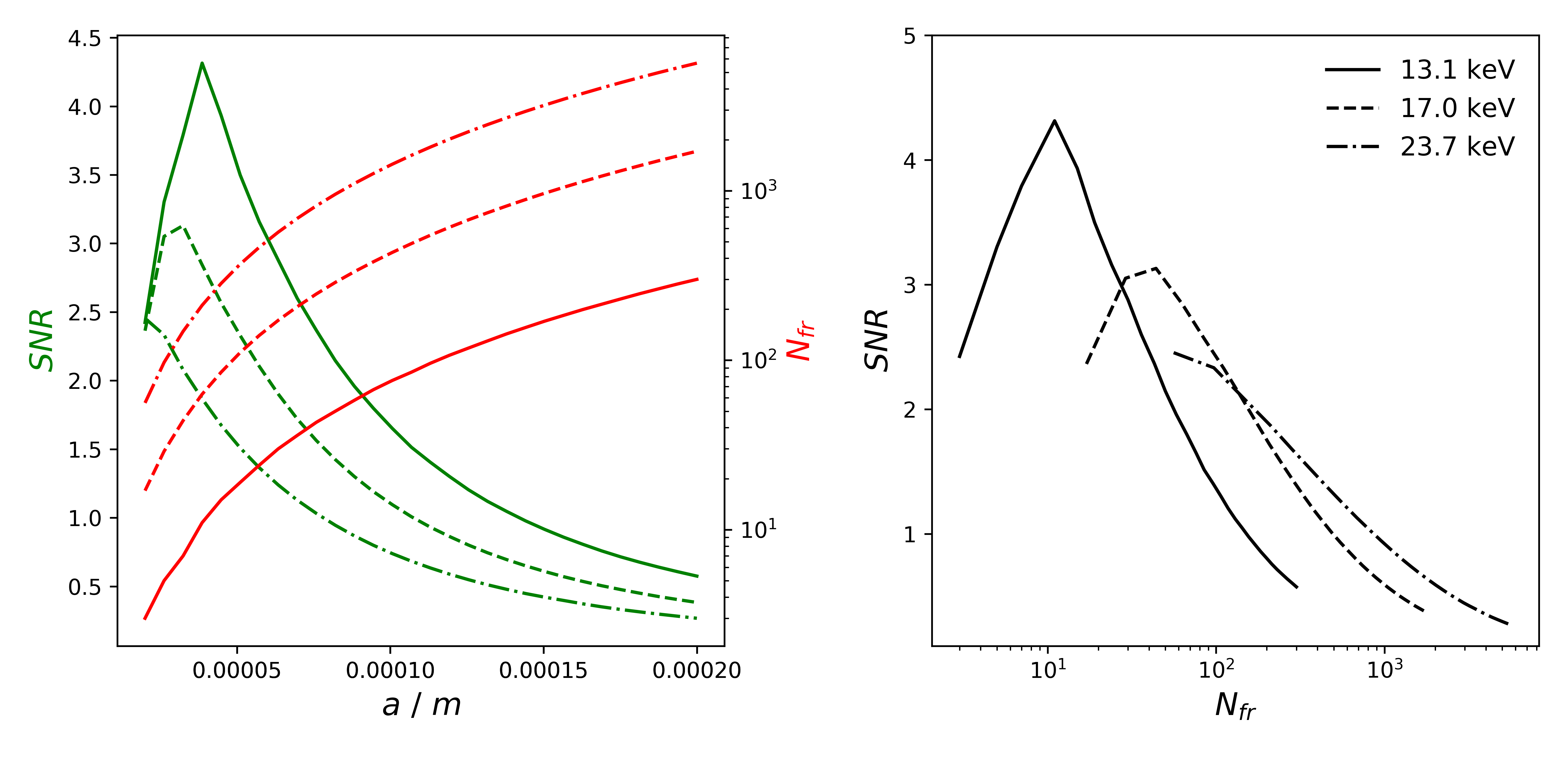}
	\caption{Left: SNR (green lines) as a function of beamsize a for photon energies of 13.1, 14.9 and 17.0 keV (solid, dashed, dashed-dotted lines). Red lines indicates the maximum number of possible frames. Right: SNR as a function of the maximum number of frames displayed for the same photon energies. }
	\label{fig:multiframe}
\end{figure}

\section{Conclusion}
We determined the signal to noise ratios (SNR) for XPCS experiments of a concentrated lysozyme solution at length scales of the hydrodynamic radius of a single protein molecule. The results show that the SNR values can at least increase up to one order of magnitude at future upgraded storage rings when compared to existing facilties. With this, the required measuring time would reduce by two orders of magnitude making dynamic studies of protein solutions at nanometer lengthscales feasible. However, in order to take full advantage of the properties of the future sources, XPCS experiments require adapted experimental setups with larger beamsizes and longer sample-detector distances than usually available at standard XPCS beamlines. 

\section{Acknowledgements}
The authors would like to thank the PETRA IV project team, and especially C.
Schroer, M. Tischer, and S. Klumpp for useful discussions and support. C.G. acknowledges funding by BMBF via project 05K19PS1.





\printbibliography
\end{document}